\documentclass{elsart}

 \usepackage{graphicx}

\usepackage{amssymb}

\begin{document}

\begin{frontmatter}



\title{The 't Hooft vertex revisited}


\author{Michael Creutz}

\address{Physics Department, Brookhaven National Laboratory\\
Upton, NY 11973, USA}

\begin{abstract}
In 1976 't Hooft introduced an elegant approach towards understanding
the physical consequences of the topological structures that appear in
non-Abelian gauge theories.  These effects are concisely summarized in
terms of an effective multi-fermion interaction.  These old arguments
provide a link between a variety of recent and sometimes controversial
ideas including discrete chiral symmetries appearing in some models
for unification, ambiguities in the definition of quark masses, and
flaws with some simulation algorithms in lattice gauge theory.
\end{abstract}

\end{frontmatter}

\section{Introduction}
\label{introduction}
More than 30 years ago 't Hooft \cite{'t Hooft:1976up,'tHooft:fv}
explored some of the physical consequences of topological structures
\cite{Belavin:1975fg} in non-Abelian gauge theories.  The issues are
directly tied to chiral anomalies, and the phenomena discussed ranged
from the mass of the $\eta^\prime$ meson to the existence of baryon
decay in the standard model.  The latter effect is too small for
observation; nevertheless, the fact that it must exist is crucial for
any fundamental formulation of the theory.  In particular, it must
appear in any valid attempt to formulate the standard model on the
lattice \cite{Eichten:1985ft}.

Although the underpinnings of these ideas have been established for
some time, recent controversies strikingly show that the issues are
not fully understood.  For example, the rooting algorithm used to
adjust the number of quark flavors in the staggered fermion lattice
algorithm is inconsistent with the expected form of the 't Hooft
vertex \cite{Creutz:2007rk}.  This has led to a rather bitter
controversy involving a large subset of the lattice gauge community
\cite{ Creutz:2007rk, Creutz:2007yg, Bernard:2006vv, Creutz:2007pr,
Kronfeld:2007ek,Bernard:2007eh}.  

A second dispute involves the speculation that a vanishing up quark
mass might solve the strong CP problem.  The 't Hooft vertex gives
rise to non-perturbative contributions to the renormalization group
flow of quark masses \cite{Georgi:1981be,Banks:1994yg}.  When the
quark masses are non-degenerate, these involve an additive shift and
show that the vanishing of a single quark mass is renormalization
scheme dependent.  As such it can not be a fundamental concept
\cite{Creutz:2003xc}.  This conflicts with the conventional
perturbative arguments that the renormalization of fermion masses is
purely multiplicative, something that is only true for multiple
degenerate flavors.  Nevertheless, various attempts to go beyond the
standard model often continue to attempt to build in a vanishing
up-quark mass as an escape from the strong CP problem; for a few
examples see
\cite{Srednicki:2005wc,Davoudiasl:2007zx,Davoudiasl:2005ai}.

All these issues are closely tied to quantum anomalies and axial
symmetries.  Indeed, when expressed in terms of the 't Hooft
interaction, the qualitative resolution of most of these effects
becomes fairly obvious.  The fact that lingering controversies
continue suggests that it is worthwhile to revisit the underpinnings
of the mechanism.  That is the purpose of this paper.  Although the
mechanism applies also to the weak interactions through the predicted
baryon decay, here I will restrict my discussion to the strong
interactions of quarks and gluons.

No single item in this discussion is new in and of itself.  The main
goal of this paper is to elucidate their unification through the 't
Hooft interaction.  I will occasionally use lattice language for
convenience, such as referring to an ultraviolet cutoff $a$ as a
``lattice spacing.''  Nevertheless, the issues are in not specific to
lattice gauge theory.  The topic is basic non-perturbative issues
within the standard quark confining dynamics of the strong
interactions.  I will rely heavily on chiral symmetries, and only
assume that I have a well regulated theory that maintains these
symmetries to a good approximation.

I organize the discussion as follows.  Section \ref{vertex} starts
with a review of how the 't Hooft effective interaction arises and
discusses some of its general properties.  In section \ref{etamass} I
turn to the historically most significant use of the effect, the
connection to the $\eta^\prime$ mass.  The robustness of the zero
modes responsible for the vertex is discussed in section \ref{index}.
The remainder of the paper goes through a variety of other physical
consequences that are perhaps somewhat less familiar.  Section
\ref{reps} explores the discrete chiral symmetries that appear with
quarks in higher representations than the fundamental, as motivated by
some unified models.  Section \ref{theta} discusses how the effective
vertex is tied to the well known possibility of CP violation in the
strong interactions through a non-trivial phase in the quark mass
matrix.  Section \ref{mu} connects the vertex to the ill posed nature
of proposing a vanishing up quark mass to solve the strong CP problem.
Building on this, section \ref{kaplanmanohar} relates this result to
the effective chiral Lagrangian ambiguity discussed by Kaplan and
Manohar \cite{Kaplan:1986ru}.  In Section \ref{axions} I briefly
discuss the axion solution to the strong CP problem, noting that the
axion does acquire a mass from the anomaly but observing that as long
as the coupling of the axion to the strong interactions is small this
mass along with inherited CP violating effects will naturally be
small.  In section \ref{rooting} I discuss why the rooting procedure
used in lattice gauge theory with staggered quarks mutilates the
interaction, thus introducing an uncontrolled approximation.  Finally,
the basic conclusions are summarized in section \ref{summary}.

\section{The vertex}
\label{vertex}

I begin with a brief reminder of the strategy of lattice simulations.
Consider the basic path integral, or ``partition function,'' for
quarks and gluons
\begin{equation}
Z=\int (dA)(d\psi)(d\overline\psi) \exp(-S_g(A)-\overline\psi D(a)\psi). 
\end{equation}
Here $A$ denotes the gauge fields and $\overline\psi,\psi$ the quark
fields.  The pure gauge part of the action is $S_g(A)$ and the matrix
describing the fermion part of the action is $D(A)$.  Since direct
numerical evaluation of the fermionic integrals appears to be
impractical, the Grassmann integrals are conventionally evaluated
analytically, reducing the partition function to
\begin{equation}
Z=\int (dA)\ e^{-S_g(A)}\ |D(A)|.
\end{equation}
Here $|D(A)|$ denotes the determinant of the Dirac matrix evaluated in
the given gauge field.  Thus motivated, the basic lattice strategy is
to generate a set of random gauge configurations weighted by
$\exp(-S_g(A))\ |D(a)|$.  Given an ensemble of such configurations,
one then estimates physical observables by averages over this
ensemble.

This procedure seems innocent enough, but it can run into trouble when
one has massless fermions and corresponding chiral symmetries.  To see
the issue, write the determinant as a product of the eigenvalues
$\lambda_i$ of the matrix $D$.  In general $D$ may not be a normal
matrix; so, one should pick either left or right eigenvectors at one's
discretion.  This is a technical detail that will not play any further
role here.  In order to control infrared issues with massless quarks,
let me introduce a small explicit mass $m$ and reduce the path
integral to
\begin{equation}
Z=\int (dA)\ e^{-S_g(A)}\ \prod (\lambda_i+m). 
\end{equation}

Now suppose we have a configuration where one of the eigenvalues of
$D(A)$ vanishes, {\it i.e.} assume that some $\lambda_i=0$.  As we
take the mass to zero, any configurations involving such an eigenvalue
will drop out of the ensemble.  At first one might suspect this would
be a set of measure zero in the space of all possible gauge fields.
However, as discussed later, the index theorem ties gauge field
topology to such zero modes of the Dirac operator.  This shows that
such modes can be robust under small deformations of the fields.
Under the traditional lattice strategy these configurations have zero
weight in the massless limit.  The naive conclusion is that such
configurations are irrelevant to physics in the chiral limit.

It was this reasoning that 't Hooft showed to be incorrect.  Indeed,
he demonstrated that it is natural for some observables to have $1/m$
factors when zero modes are present.  These can cancel the terms
linear in $m$ from the determinant, leaving a finite contribution.

As a simple example, consider the quark condensate 
\begin{equation}
\langle \overline\psi \psi\rangle=
{1\over VZ}
\int (dA)\ 
e^{-S_g}\ |D|\ \ {\rm Tr}D^{-1}.
\end{equation}
Here $V$ represents the system volume, inserted to give an intensive
quantity.  Expressing the fermionic factors in terms of the
eigenvalues of $D$ reduces this to
\begin{equation}
\langle \overline\psi \psi\rangle=
{1\over VZ}
\int (dA)\ 
e^{-S_g}\ 
\prod (\lambda_i+m)
\ \ \sum_i {1\over \lambda_i+m}.
\end{equation}
Now if there is a mode with $\lambda_i=0$, the factor of $m$ is
canceled by a $1/m$ piece in the trace of $D^{-1}$.  Configurations
containing a zero mode give a constant contribution to the condensate
that survives in the chiral limit.  Note that this effect is unrelated
to spontaneous breaking of chiral symmetry and appears even with
finite volume.

This contribution to the condensate is special to the one-flavor
theory.  Because of the anomaly, this quark condensate is not an order
parameter for any symmetry.  With more fermion species there will be
additional factors of $m$ from the determinant.  Then the effect is of
higher order in the fermion fields and does not appear directly in the
condensate.  For two or more flavors the standard Banks-Casher picture
\cite{Banks:1979yr} of an eigenvalue accumulation leading to the
spontaneous breaking of chiral symmetry should apply.

The conventional discussion of the 't Hooft vertex starts by inserting
fermionic sources into the path integral
\begin{equation}
Z(\eta,\overline\eta)=\int (dA)\ (d\psi)\ (d\overline\psi)\
e^{-S_g-\overline\psi (D+m) \psi +\overline\psi \eta+
\overline\eta\psi}.
\end{equation}
Differentiation, in the Grasmannian sense, with respect to these
sources can generate the expectation for an arbitrary product of
fermionic operators.  Integrating out the fermions reduces this to
\begin{equation}
Z=\int (dA)\ 
e^{-S_g{+\overline\eta(D+m)^{-1}\eta}}\ 
\prod (\lambda_i+m). 
\end{equation}
Consider a zero mode $\psi_0$ satisfying {$D\psi_0=0$}.  If the source
has an overlap with the mode, that is $(\psi_0^\dagger\cdot\eta)\ne
0$, then a factor of {$1/m$} in the source term can cancel the {$m$}
from the determinant.  Although non-trivial topological configurations
do not contribute to $Z$, their effects can survive in correlation
functions.  For the one-flavor theory the effective interaction is
bilinear in the fermion sources and is proportional to
\begin{equation}
(\overline\eta\cdot\psi_0)(\psi_0^\dagger\cdot\eta).
\label{bilinear}
\end{equation}
As discussed later, the index theorem tells us that in general the
zero mode is chiral; it appears in either {$\overline\eta_L\eta_R$} or
{$\overline\eta_R\eta_L$}, depending on the sign of the gauge field
winding.

\begin{figure*}
\centering
\includegraphics[width=2.5in]{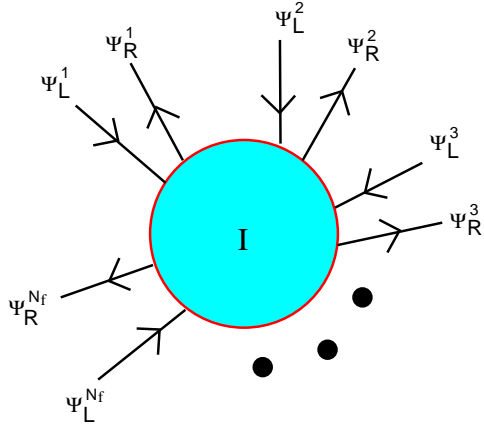}
\caption{ The 't Hooft vertex for $N_f$ flavors is a $2N_f$ effective
fermion operator that flips the spin of every flavor.}
\label{instanton} 
\end{figure*}

With $N_f\ge2$ flavors, the cancellation of the mass factors in the
determinant requires source factors from each flavor.  This
combination is the 't Hooft vertex.  It is an effective $2N_f$ fermion
operator.  In the process, every flavor flips its spin, as sketched in
Fig.~\ref{instanton}.  Indeed, this is the chiral anomaly; left and
right helicities are not separately conserved.

Because of Pauli statistics, the multi-flavor vertex can be written in
the form of a determinant.  This clarifies how the vertex preserves
flavored chiral symmetries.  With two flavors, call them $u$ and $d$,
Eq.~\ref{bilinear} generalizes to
\begin{equation}
\left\vert
\matrix{
(\overline u \cdot\psi_0)(\psi_0^\dagger\cdot u) 
& (\overline u \cdot\psi_0)(\psi_0^\dagger\cdot d)\cr
(\overline d \cdot\psi_0)(\psi_0^\dagger\cdot u) 
& (\overline d \cdot\psi_0)(\psi_0^\dagger\cdot d)\cr
}
\right\vert.
\end{equation}

Note that the effect of the vertex is non-local.  In general the zero
mode $\psi_0$ is spread out over some finite region.  This means there
is an inherent position space uncertainty on where the fermions are
interacting.  A particular consequence is that fermion conservation is
only a global symmetry.  In Minkowski space language, this
non-locality can be thought of in terms of states sliding in and out
of the Dirac sea at different locations.

\section{ The $\eta^\prime$ mass}
\label{etamass}

The best known consequence of the 't Hooft interaction is the
explanation of why the $\eta^\prime$ meson is substantially heavier
than the other pseudo-scalars.  Consider the three-flavor theory with
up, down, and strange quarks, $u,d,s$.  The quark model indicates that
this theory should have three neutral non-strange pseudo-scalars.
Experimentally these are the $\pi_0$ at 135 MeV, the $\eta$ at 548
Mev, and the $\eta^\prime$ at 958 MeV.  In the quark model, these
should be combinations of the quark bound states $\overline u\gamma_5
u,\ \overline d\gamma_5 d,\ \overline s\gamma_5 s$.

In the standard chiral picture, the squares of the Goldstone boson
masses are linear in quark masses.  The strange quark is the heaviest
of the three, with its mass related to $m_K=$ 498 MeV.  The maximum
mass a Goldstone boson could have would be if it is pure $\overline s
s$.  Ignoring the light quark masses, this maximum value is $\sqrt 2
m_K = 704\ \hbox{MeV}$, substantially less than the observed mass of
the ${\eta^\prime}$.  From this we are driven to conclude that the
$\eta^\prime$ must be something else and not a Goldstone boson.

When viewed in the context of the 't Hooft interaction, the problem
disappears.  The vertex directly breaks the naive $U(1)$ axial
symmetry, and thus there is no need for a corresponding Goldstone
boson.  Thus, the mass of the $\eta^\prime$ should be of order the
strong interaction scale plus the masses of the contained quarks.
Indeed, when compared to a vector meson mass, such as the $\phi$ at
1019 MeV, the 958 MeV of the $\eta^\prime$ seems quite normal.

Even though this resolves the issue, it is perhaps interesting to look
a bit further into the differences between the singlet and the
flavored pseudo-scalar mesons.  For the one-flavor case there are two
effects that give extra contributions to the singlet mass.  First, the
vertex itself gives a direct mass shift to the quark, and, second, the
vertex directly couples the quark-antiquark content to gluonic
intermediate states.  In general it is expected that
\begin{equation}
\langle \eta^\prime|F\tilde F|0\rangle\ne 0.
\end{equation}
The $\eta^\prime$ can be created not just by quark operators but also
a pseudo-scalar gluonic combination.

\begin{figure*}
\centering
\includegraphics[width=3in]{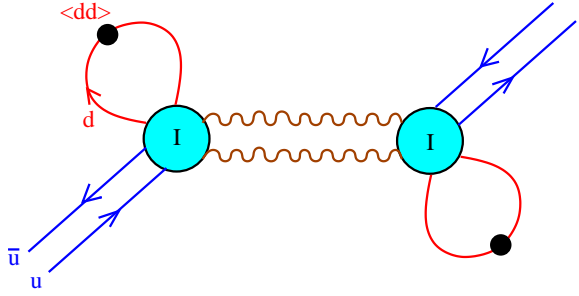}
\caption{The 't Hooft vertex couples the quark-antiquark content of
the pseudo-scalar meson with gluonic intermediate states.  Additional
quark lines associated with the vertex can be absorbed in the
condensate.}
\label{etamassfig} 
\end{figure*}

With more quark species, flavored pseudo-scalar Goldstone bosons
should exist.  Their primary difference from the singlet is the
absence of the gluonic intermediate states.  It is the 't Hooft vertex
that non-locally couples $F\tilde F$ to the quark-antiquark content of
the flavor singlet meson.  Note that with multiple flavors the vertex
involves more than two fermion lines; the contributions of the extra
lines can be absorbed in the condensate, as sketched in
Fig.~\ref{etamassfig}.  This large mass generation is sometimes
referred to as coming from ``constituent'' quark masses, as opposed to
the ``current'' quark masses that vanish in the chiral limit.

The renormalization group provides useful information on the coupling
constant dependence of the $\eta^\prime$ mass as the cutoff is
removed.  These equations read
\begin{equation}
a{dg\over da}=\beta(g)=\beta_0 g^3+\beta_1 g^5 +\ldots
+{\rm non{\hbox{-}}perturbative}
\end{equation}
for the bare coupling constant $g$ and 
\begin{equation}
a{dm\over da}=m\gamma(g)=m(\gamma_0 g^2+\gamma_1 g^4 +\ldots)
+{\rm non{\hbox{-}}perturbative}
\label{mrg}
\end{equation}
for the bare quark mass $m$.  I include this latter equation for later
discussion.  As is well known, the coefficients $\beta_0$, $\beta_1$,
and $\gamma_0$ are independent of renormalization scheme.  It is
important to remember that the separation of the perturbative and
non-perturbative parts of the renormalization group equations is
scheme dependent.  Indeed, different definitions of the coupling
constant will in general differ by non-perturbative parts.  This will
play an important role later when I discuss non-perturbative changes
in the definition of quark masses.

Renormalizing by holding the $\eta^\prime$ mass fixed allows the
solution of the the coupling constant equation with the result
\begin{equation}
m_{\eta^\prime}= 
C\ { e^{-1/2\beta_0 g^2} g^{-\beta_1/\beta_0^2}\over a}
\times(1+O(g^2))=O( \Lambda_{qcd}).
\label{mofa}
\end{equation}
Here $C$ is a dynamically determined constant which could in principle
be determined in numerical simulations.  Because the inverse of the
coupling appears in the exponent, this dependence is non-perturbative.
Indeed, this equation is a simple restatement of asymptotic freedom,
the requirement that $\lim_{a\rightarrow 0}\ g(a)=0$ logarithmically.
The importance of this relation is that similar expressions involving
exponential dependences on the inverse coupling are natural and
expected to occur in any quantities where non-perturbative effects
are important.

\section{Robustness and the index theorem}
\label{index}

The reason these zero modes remain crucial is their robustness through
the connection to the index theorem \cite{Atiyah:1971rm}.  Otherwise
they could be argued to contribute a set of measure zero to the path
integral.  When the gauge field is smooth, then the difference in the
number of right handed and left handed zero modes is tied to a
topological wrapping of the gauge fields at infinity around the gauge
group.  Being topological, this winding is robust under small
deformations of the gauge fields.  Therefore exact zero modes are not
accidental but required whenever gauge configurations have non-trivial
topology.

The robustness of these zero modes can also be seen directly from the
eigenvalue structure of the Dirac operator.  This builds on
$\gamma_5$ hermiticity
\begin{equation}
D^\dagger=\gamma_5 D\gamma_5,
\end{equation}
a condition true in the naive continuum theory as well as most lattice
discretizations.  A direct consequence is that non-real eigenvalues of
$D$ occur in complex conjugate pairs.  All eigenvalues $\lambda$
satisfy $|D-\lambda|=0$, where the vertical lines denote the
determinant.  This plus the fact that $|\gamma_5|=1$ gives
\begin{equation}
|D-\lambda^*|=|\gamma_5 (D^\dagger-\lambda^*)\gamma_5|=
|D^\dagger-\lambda^*|=|D-\lambda|^*=0. 
\end{equation}
Thus all eigenvalues are either in complex pairs or real.  

Close to the massless continuum limit, $D$ is predominantly
anti-Hermitian.  Small real eigenvalues correspond to the zero modes
that generate the 't Hooft vertex.  Ignoring possible lattice
artifacts, $D$ should approximately anticommute with $\gamma_5$.  If
$D\psi=0$, then $D\gamma_5\psi=-\gamma_5 D\psi=0.$ Thus $D$ and
$\gamma_5$ commute on the subspace spanned by all the zero modes.
Restricted to this space, these matrices can be simultaneously
diagonalized.  Since all eigenvalues of $\gamma_5$ are plus or minus
unity, the trace of $\gamma_5$ restricted to this subspace must be an
integer.  Because of this quantization, this integer will generically
be robust under small variations of the gauge fields.  This defines
the index for the given gauge field.

This approach of defining the index directly through the eigenvalues
of the Dirac operator has the advantage over the topological
definition in that the gauge fields need not be differentiable.  For
smooth fields the definitions are equivalent through the index
theorem.  But in general path integrals are dominated by
non-differentiable fields.  Also, on the lattice the gauge fields lose
the precise notion of continuity and topology.

In the fermion approach to the index other subtleties do arise.  In
general a regularization can introduce distortions so that the real
eigenvalues are not necessarily exactly at the same place.  In
particular, for Wilson lattice fermions the real eigenvalues spread
over a finite range.  In addition to the small eigenvalues near zero,
the Wilson approach has additional real eigenvalues far from the
origin that are associated with doublers.  Applying $\gamma_5$ to a
small eigenvector will in general mix in a small about of the larger
modes.  This allows the trace of $\gamma_5$ on the subspace involving
only the low modes to deviate from an exact integer.

The overlap operator \cite{Neuberger:1997fp} does constrain the small
real eigenvalues to be at the origin and the earlier argument goes
through.  In this case additional real eigenvalues do occur far from
the origin.  These are required so that the trace of $\gamma_5$ over
the full space will vanish.  Since the overlap operator keeps the
index discrete, it is forced to exhibit discontinuous behavior as the
gauge fields vary between topological sectors.  In the vicinity of
these discontinuities the gauge fields can be thought of as ``rough''
and the precise value of the index can depend on the details of the
kernel used to project onto the overlap matrix.  For multiple light
fermion flavors this ambiguity in the index is expected to be
suppressed in the continuum limit.  Nevertheless, the issues discussed
later for a single massless quark suggests that the situation may be
more subtle for the zero or one species case.

\section{Fermions in higher representations of the gauge group}
\label{reps}

When the quarks are massless, the classical field theory corresponding
to the strong interactions has a $U(1)$ axial symmetry under the
transformation
\begin{equation} 
\psi\rightarrow
e^{i\theta\gamma_5}\psi\qquad
\overline\psi\rightarrow \overline\psi e^{i\theta\gamma_5}.
\label{thetarot}
\end{equation}
It is the 't Hooft vertex that explains how this symmetry does not
survive quantization.  In this section I discuss how in some special
cases, in particular when the quarks are in non-fundamental
representations of the gauge group, discrete subgroups of this
symmetry can remain.

While these considerations do not apply to the usual theory of the
strong interactions, there are several reasons to study them anyway.
At higher energies, perhaps as will be probed at the upcoming Large
Hadron Collider, one might well discover new strong interactions that
play a substantial role in the spontaneous breaking of the electroweak
theory.  Also, many grand unified theories involve fermions in
non-fundamental representations.  As one example, massless fermions in
the 10 representation of $SU(5)$ possess a $Z_3$ discrete chiral
symmetry.  Similarly the left handed 16 covering representation of
$SO(10)$ gives a chiral gauge theory with a surviving discrete $Z_2$
chiral symmetry.  Understanding these symmetries may play some role in
an eventual discretization of chiral gauge theories on the lattice.

I build here on generalizations of the index theorem relating gauge
field topology to zero modes of the Dirac operator.  In particular,
fermions in higher representations can involve in multiple zero modes
for a given winding.  Being generic, consider representation $X$ of a
gauge group $G$.  Denote by $N_X$ the number of zero modes that are
required per unit of winding number in the gauge fields.  That is,
suppose the index theorem generalizes to
\begin{equation}
n_r-n_l=N_X\nu
\end{equation}
where $n_r$ and $n_l$ are the number of right and left handed zero
modes, respectively, and $\nu$ is the winding number of the associated
gauge field.  The basic 't Hooft vertex receives contributions from
each zero mode, resulting in an effective operator which is a product
of $2N_X$ fermion fields.  Schematically, the vertex is modified along
the lines $\overline\psi_L \psi_R \longrightarrow (\overline\psi_L
\psi_R)^{N_X}$.  While this form still breaks the $U(1)$ axial
symmetry, it is invariant under $\psi_R\rightarrow e^{2\pi
i/N_X}\psi_R$.  In other words, there is a $Z_{N_x}$ discrete axial
symmetry.

There are a variety of convenient tools for determining $N_X$.
Consider building up representations from lower ones.  Take two
representations $X_1$ and $X_2$ and form the direct product
representation $X_1\otimes X_2$.  Let the matrix dimensions for $X_1$
and $X_2$ be $D_1$ and $D_2$, respectively.  Then for the product
representation we have
\begin{equation}
N_{X_1\otimes X_2}= N_{X_1} D_{X_2}+N_{X_2} D_{X_1}.
\end{equation}
To see this, start with $X_1$ and $X_2$ representing two independent
groups $G_1$ and $G_2$.  With $G_1$ having winding, there will be a
zero mode for each of the dimensions of the matrix index associated
with $X_2$.  Similarly there will be multiple modes for winding in
$G_2$.  These modes are robust and all should remain if we now
constrain the groups to be the same.

As a first example, denote the fundamental representation of $SU(N)$
as $F$ and the adjoint representation as $A$.  Then using $\overline F
\otimes F = A+1$ in the above gives $N_A=2N$, as noted some time ago
in Ref.~\cite{Witten:1982df}.  With $SU(3)$, fermions in the adjoint
representation will have six-fold degenerate zero modes.

For another example, consider $SU(2)$ and build up towards arbitrary
spin $s\in\{0,{1\over 2}, 1, {3\over 2},\ldots\}$.  Recursing the
above relation gives the result for arbitrary spin
\begin{equation}
N_s=s(2s+1)(2s+2)/3.
\end{equation}

Another technique for finding $N_X$ in more complicated groups begins
by rotating all topological structure into an $SU(2)$ subgroup and
then counting the corresponding $SU(2)$ representations making up the
larger representation of the whole group.  An example to illustrate
this procedure is the antisymmetric two indexed representation of
$SU(N)$.  This representation has been extensively used in
\cite{Corrigan:1979xf,Armoni:2003fb,Sannino:2003xe,Unsal:2006pj} for
an alternative approach to the large $N_c$ limit.  The basic
$N(N-1)/2$ fermion fields take the form
\begin{equation}
\psi_{ab}=-\psi_{ba}, \qquad a,b\in 1,2,...N.
\end{equation}
Consider rotating all topology into the $SU(2)$ subgroup involving the
first two indices, i.e. 1 and 2.  Because of the anti-symmetrization,
the field $\psi_{12}$ is a singlet in this subgroup.  The field pairs
$(\psi_{1,j},\psi_{2,j})$ form a doublet for each $j\ge 3$.  Finally,
the $(N-2)(N-3)/2$ remaining fields do not transform under this
subgroup and are singlets.  Overall we have $N-2$ doublets under the
$SU(2)$ subgroup, each of which gives one zero mode per winding
number.  We conclude that the 't Hooft vertex leaves behind a
$Z_{N-2}$ discrete chiral symmetry.  Specializing to the 10
representation of $SU(5)$, this is the $Z_3$ mentioned earlier.

Another example is the group $SO(10)$ with fermions in the 16
dimensional covering group.  This forms the basis of a rather
interesting grand unified theory, where one generation of fermions is
placed into a single left handed 16 multiplet \cite{Georgi:1979dq}.
This representation includes two quark species interacting with the
$SU(3)$ subgroup of the strong interactions, Rotating a topological
excitation into this subgroup, we see that the effective vertex will
be a four fermion operator and preserve a $Z_2$ discrete chiral
symmetry.

\begin{figure*}
\centering
\includegraphics[width=2.5in]{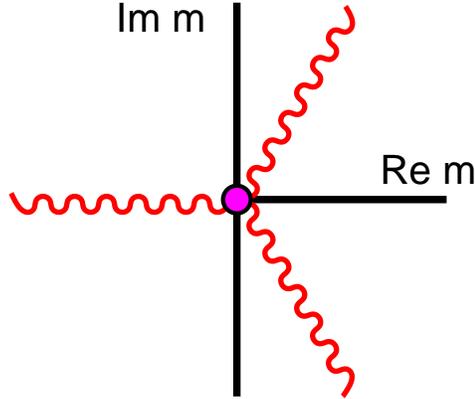}
\caption{With massless fermions in the 10 representation of gauge
group $SU(10)$ there exists a discrete $Z_3$ chiral symmetry.  If this
is spontaneously broken one expects three phase transitions to meet at
the origin in complex mass space, as sketched here. (From
Ref.~\cite{Creutz:2006ts}).}
\label{sufivealt} 
\end{figure*}

\begin{figure*}
\centering
\includegraphics[width=2.5in]{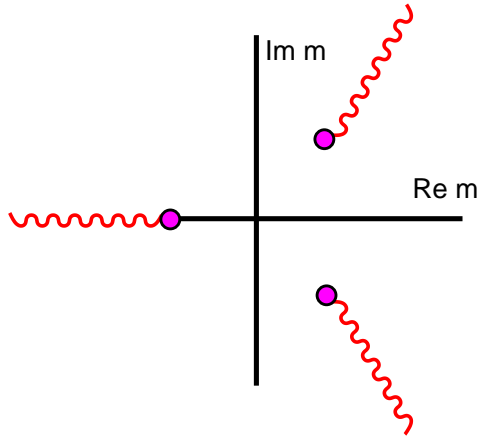}
\caption{If the discrete chiral symmetry is not broken spontaneously,
$SU(5)$ gauge theory with fermions in the 10 representation should
behave smoothly in the quark mass as it passes through zero.  Such a
smooth behavior is similar to that expected for the one-flavor theory
in the fundamental representation.  (From Ref.~\cite{Creutz:2006ts}).}
\label{sufive} 
\end{figure*}

It is unclear whether these discrete symmetries are expected to be
spontaneously broken.  Since they are discrete, such breaking is not
associated with Goldstone bosons.  But the quark condensate does
provide an order parameter; so when $N_X>1$, any such breaking would
be conceptually meaningful.  Returning to the $SU(5)$ case with
fermions in the 10, a spontaneous breaking would give rise to discrete
jumps in this order parameter as a function of the complex mass plane,
as sketched in Fig.~\ref{sufivealt}.  Alternatively, the unbroken
theory would have a phase diagram more like that in Fig.~\ref{sufive}.
In these figures I assume that for large mass a spontaneous breaking
of parity does occur when the strong CP violation angle is set to
$\pi$.  Such a jump is expected even for the one-flavor theory with
fermions in the fundamental representation \cite{Creutz:2006ts}.

Which of these behaviors is correct could be determined in lattice
simulations, although there are issues in how the lattice formulation
is set up.  The Wilson approach involves irrelevant chiral symmetry
breaking operators that will in general distort the three fold
symmetry of these models.  Even the overlap operator
\cite{Neuberger:1997fp}, which respects a variation of the continuous
chiral symmetries, appears to break these discrete symmetries
\cite{Edwards:1998dj}.  Nevertheless, as one comes sufficiently close
to the continuum limit, it should be possible to distinguish between
these scenarios.

\section{The Theta parameter and the 't Hooft vertex}
\label{theta}

When the quarks are massless, the classical field theory corresponding
to the strong interactions has a $U(1)$ axial symmetry under the
transformation in Eq.~(\ref{thetarot}).  On the other hand, a fermion
mass term, say $m\overline\psi\psi$, breaks this symmetry explicitly.
Indeed, under the chiral rotation of Eq.~(\ref{thetarot})
\begin{equation}
m\overline\psi\psi\rightarrow m\cos(\theta)\overline\psi\psi
+im\sin(\theta)\overline\psi\gamma_5\psi
\label{m5}
\end{equation}
If the classical chiral symmetry of the kinetic term was not broken by
quantum effects, then a mass term of the form of the right hand side
of this equation would be physically completely equivalent to the
normal mass term.  But because of the effect of the 't Hooft
interaction, the theory with the rotated mass is physically
inequivalent to the unrotated theory.

However the theory is regulated, it is essential that the cutoff
distinguish between the two terms on the right hand side of
Eq.~(\ref{m5}).  With a Pauli-Villars scheme it is the the mass for
the heavy regulator field that fixes the angle $\theta$.  For Wilson
fermions the Wilson term selects the chiral direction.  This carries
over to the overlap formulation, built on a projection from the Wilson
operator.  Unfortunately it is the absence of such a distinction that
lies at the heart of the failure of the rooting prescription for
staggered fermions, as discussed later.

The above rotation is often described by complexifying the mass
term.  If we write
\begin{equation}
\overline\psi\psi=\overline\psi_L\psi_R+\overline\psi_R\psi_L
\end{equation}
with 
\begin{eqnarray}
\psi_{R,L}={1\pm\gamma_5\over 2}\ \psi\\
\overline\psi_{R,L}=\overline\psi\ {1\mp\gamma_5\over 2},
\end{eqnarray}
then our generalized mass term takes the form
$$
m\overline\psi_L\psi_R+m^*\overline\psi_R\psi_L
$$ with $m=|m|e^{i\theta}$ a complex number.  In this latter notation,
the effect of the 't Hooft vertex is to make the phase of the mass
matrix an observable quantity.  This phase is connected to the strong
$CP$ angle, usually called $\Theta$.

Indeed, because of this effect, the real and the imaginary parts of
the quark masses are actually independent parameters.  The two terms
$\overline\psi\psi$ and $i\overline\psi\gamma_5\psi$, which are
naively equivalent, are in fact distinct possible ways to break the
chiral symmetry.  It is the 't Hooft vertex which distinguishes one of
them a special.  With usual conventions $i\overline\psi\gamma_5\psi$
is a CP odd operator; therefore, its interference with the vertex can
generate explicit CP violation.  The non-observation of such in the
strong interactions indicates this term must be quite small; this lies
at the heart of the strong CP problem.

With multiple flavors the possibility of flavored axial chiral
rotations allows one to move the phase of the mass between the various
species without changing the physical consequences.  One natural
choice is to place all phases on the lightest quark, say the up quark,
and keep all others real.  Equivalently one could put all the phase on
the top quark, but this would obscure the effects on low energy
physics.  If one gives all quarks a common phase $\theta$, then that
phase is is related to the physical parameter by $\theta=\Theta/N_f$.

\section{The strong CP problem and $m_u=0$} 
\label{mu}

One of the puzzles of the strong interactions is the experimental
absence of CP violation, which would not be the case if the imaginary
part of the mass were present.  This would be quite unnatural if at
some higher energy the strong interactions were unified with the weak
interactions, which are well known not to satisfy CP symmetry.  On
considering the strong interactions at lower energies, some residual
effect of this breaking would naturally appear in the basic
parameters, in particular through the imaginary part of the quark
mass.  The apparent experimental absence of such is known as the
strong CP puzzle.

An old suggestion to resolve this puzzle is that one of the quark
masses might vanish.  Indeed, this is a bit of a tautology since if it
vanishes as a complex parameter, so does its imaginary part.  But the
imaginary part is really an independent parameter, and so it seems
quite peculiar to tie it to the real part.  While phenomenological
models suggest that the up quark mass is in fact far from vanishing,
various attempts to go beyond the standard model continue to attempt
building in a vanishing up-quark mass at some high scale as an escape
from the strong CP problem
\cite{Srednicki:2005wc,Davoudiasl:2007zx,Davoudiasl:2005ai}.

It is through consideration of the 't Hooft vertex that one sees that
this solution is in fact ill posed \cite{Creutz:2003xc}.  As discussed
earlier, for the one-flavor case the vertex introduces a shift in the
quark mass of order $\Lambda_{qcd}$.  The amount of this shift will in
general depend on the details of the renormalization group scheme and
the scale of definition.  The concept of a vanishing mass is not a
renormalization group invariant, and as such it should not be relevant
to a fundamental issue such as whether the strong interactions violate
CP symmetry.

\begin{figure*}
\centering
\includegraphics[width=2.5in]{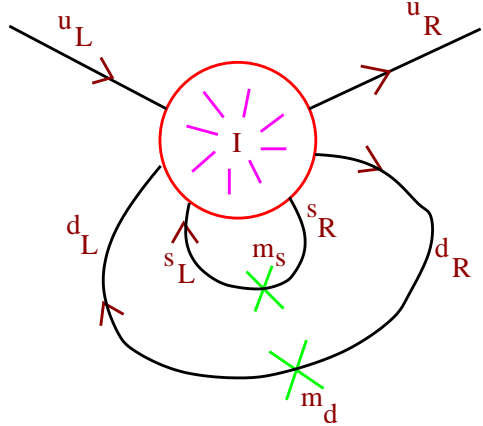}
\caption{With three non-degenerate flavors the lines representing the
heavier quarks can be joined to 't Hooft vertex in such a way that the
combination gives an ambiguity in the light quark mass of order the
product of the heavier masses. (From
Ref.~\cite{Creutz:2003xc}).}
\label{thooft} 
\end{figure*}

This point carries over into the theory with multiple flavors as long
as they are not degenerate.  The experimental fact that the pion mass
does not vanish indicates that two independent flavors cannot both be
massless.  If one considers the multiple flavor 't Hooft vertex, then
one can always absorb the involved heavy quark lines with their
masses, as sketched in Fig.~\ref{thooft}.  This leaves behind a
residual bilinear fermion vertex of order the product of the heavier
quark masses.  For the three flavor theory, this gives an ambiguity in
the definition of the up quark mass of order $m_dm_s/\Lambda_{qcd}$
\cite{Georgi:1981be,Banks:1994yg}.  Note that it is the mass
associated with chiral symmetry, i.e. the ``current'' quark mass, that
is being considered here; thus, the heavy quark lines cannot be
absorbed in the condensate as they were in the earlier discussion of
the $\eta^\prime$ mass.

Can we define a massless quark via its bare value?  This approach
fails at the outset due to the perturbative divergences inherent in
the bare parameters of any quantum field theory.  Indeed, the
renormalization group tells us that the bare quark mass must be zero,
regardless of the physical hadronic spectrum.  One immediate
consequence is that it does not make sense to take the continuum limit
before taking the mass to zero; the two limits are intricately
entwined through the renormalization group equations.

To see this explicitly, recall the renormalization group equation for
the mass, Eq.~(\ref{mrg}).  This is easily solved to reveal the small
cutoff behavior of the bare mass
\begin{equation}
m=M_R\ g^{\gamma_0/\beta_0}
(1+O(g^2)).
\end{equation}
This goes to zero as $a\rightarrow 0$ since $g$ does so by asymptotic
freedom and $\gamma_0/\beta_0>0$.  Here {$M_R$} denotes an integration
constant which might be regarded as a ``renormalized mass.''  One
cannot sensibly use $M_R$ to define a vanishing mass since it has an
additive ambiguity.  For example, consider a non-perturbative
redefinition of the bare mass
\begin{equation}
\tilde m_0=m_0- g^{\gamma_0/\beta_0}\times
{ e^{-1/2\beta_0 g^2} g^{-\beta_1/\beta_0^2}\over
a}\times {\Delta\over  \Lambda_{qcd}}.
\label{tildem}
\end{equation}
This is still a solution of the renormalization group equation, but
involves the shift
\begin{equation}
M_R\rightarrow M_R-\Delta.
\end{equation}
Since the parameter $\Delta$ can be chosen arbitrarily, a vanishing of
the renormalized mass for a non-degenerate quark is meaningless.
While the exponential factor in Eq.~(\ref{tildem}) may look contrived,
non-perturbative forms like this are in fact natural.  Indeed, compare
this expression with that for the eta prime mass, Eq.~(\ref{mofa}).

\section{Connections with the Kaplan-Manohar ambiguity}
\label{kaplanmanohar}

In 1986 Kaplan and Manohar \cite{Kaplan:1986ru}, working in the
context of next to leading order chiral Lagrangians, pointed out an
inherent ambiguity in the quark masses.  This takes a similar form to
that found above, being proportional to the product of the heavier
quark masses.  The appearance of this form from the 't Hooft vertex is
illustrative of a fundamental connection to the phenomenological
chiral Lagrangian models.

In this section I slightly rephrase the Kaplan-Manohar argument.
Consider the three flavor theory with mass matrix
\begin{equation}
M=\pmatrix{
m_u & 0 & 0\cr
0 & m_d & 0\cr
0 & 0 & m_s\cr
}
\label{massmatrix}
\end{equation}
Chiral symmetry manifests itself in the massive theory as an
invariance of physical quantities under changes in the quark mass
matrix.  Under a rotation of the form
\begin{equation}
M\rightarrow g_L M g_R^{-1}
\label{chiralm}
\end{equation}
the basic physics of particles and their scatterings will remain
equivalent.  Here $g_L$ and $g_R$ are arbitrary elements of the flavor
group, here taken as $SU(3)$.

To proceed, consider the invariance of the antisymmetric tensor
under the flavor group
\begin{equation}
 \epsilon_{abc}=g_{ac}g_{bd}g_{ce}\epsilon_{cde}.
\end{equation}
Using this, it is straightforward to show that the combination
\begin{equation}\epsilon_{acd}\epsilon_{bef}
M^\dagger_{ec}M^\dagger_{fd}
\label{chiralm1}
\end{equation}
transforms exactly the same way as $M$ under the change in
Eq.~(\ref{chiralm}).  This symmetry allows the renormalization group
equations to mix the term in Eq.~(\ref{chiralm1}) with the starting
mass matrix.  Under a change of scale, the mass matrix can evolve to a
combination along the lines
\begin{equation}
M_{ab}\rightarrow \alpha M_{ab}+\beta\epsilon_{acd}\epsilon_{bef}
M^\dagger_{ec}M^\dagger_{fd}
\end{equation}
Writing this in terms of the three quark masses in
Eq.~(\ref{massmatrix}) gives
\begin{equation}
m_u\rightarrow \alpha m_u+2 \beta m_sm_d
\end{equation}
This is exactly the same form as generated by the 't Hooft vertex.

For the three flavor theory this is a next-to-leading-order chiral
ambiguity in $m_u$.  Dropping down to less flavors the issue becomes
sharper, being a leading-order mixing of $m_u$ with $m_d^*$ in the two
flavor case.  For one flavor it is a zeroth-order effect, leaving a
mass ambiguity of order $\Lambda_{qcd}$.

Increasing the number of light flavors tends to suppress topologically
non-trivial gauge configurations.  Effectively the fermions act to
smooth out rough gauge fields.  If we drop down in the number of
flavors even further towards the pure Yang-Mills theory, the
fluctuations associated with topology should become still stronger.
Indeed, the issues present in the one-flavor case suggest that there
may be a residual ambiguity in defining topological susceptibility
for the pure glue theory \cite{Creutz:2004ir}.

Since the theta parameter arises from topological issues in the gauge
theory, one might wonder how its effects can be present in the chiral
Lagrangian approach, where the gauge fields are effectively hidden.
The reason is tied to the constraint that the effective fields are in
the group $SU(3)$ rather than $U(3)$.  The chiral Lagrangian imposes
from the outset the correct symmetry of the theory including
anomalies.  Had one worked with a $U(3)$ effective field, then one
would need to add a term to break the unwanted axial $U(1)$.
Involving the determinant of the effective matrix accomplishes this,
as in Ref.~\cite{Di Vecchia:1980ve}.

\section{Axions and the strong CP problem}
\label{axions}

Another approach to the strong CP issue is to make the imaginary part
of the quark mass a dynamical quantity that naturally relaxes to zero.
Excitations of this new dynamical field are referred to as axions, and
this is known as the axion solution to the strong CP problem.  The
basic idea is to replace a quark mass term with a coupling to a new
dynamical field ${\cal A}(x)$
\begin{equation}
m\overline\psi_L \psi_R + \hbox{h.c.}\rightarrow m \overline\psi_L \psi_R +
i\xi {\cal A}(x) \overline\psi_L \psi_R + \hbox{h.c.}
+    (\partial_\mu {\cal A}(x))^2/2
\end{equation}
Here $\xi$ is a parameter that allows one to adjust the strength of
the axion coupling to hadrons; if $\xi$ is sufficiently small, the
axion would not be observable in ordinary hadronic interactions.  Any
imaginary part in $m$ can then be shifted away, thus removing CP
violation.  This is the Peccei-Quinn symmetry \cite{Peccei:1977hh}.

At this level the axion is massless.  However, the operator coupled to
the axion field can create eta prime mesons, so this term will mix the
axion and the eta prime.  Since non-perturbative effects give the eta
prime a large mass, this mixing will in general not leave the physical
axion massless; indeed it should acquire a mass of order $\xi^2$.

This requirement for a renormalization of the axion mass shows how the
anomaly forces a breaking of the Peccei-Quinn symmetry.  As that was
motivated by the strong CP problem, one might wonder if the axion
really still solves this issue.  As long as CP violation is present in
some unified theory, and we don't have the shift symmetry, the
reduction to the strong interactions could leave behind a linear term
in the axion field, i.e. something that cannot be shifted away.  The
fact that we are taking $\xi$ small suggests that such a term would
naturally be of order $\xi^2$, i.e. something of order the axion
mixing with the eta prime.  As long as the axion mass is not large,
the visible CP violations in the strong interactions will remain small
and the axion solution to the strong CP problem remains viable.

\section{Consequences for rooted determinants}
\label{rooting}

Among the more controversial consequences of the 't Hooft vertex is
the fact that it is severely mutilated by the ``rooting trick''
popular in many lattice gauge simulations
\cite{Creutz:2007yg,Creutz:2007rk}.  This represents a serious flaw in
these algorithms.  Indeed, the main purpose of lattice gauge theory is
to obtain non-perturbative information on field theories, and the 't
Hooft interaction is one of the most important non-perturbative
effects.  Nevertheless, the large investments that have been made in
such algorithms has led some authors to attempt refuting this flaw
\cite{Bernard:2006vv,Kronfeld:2007ek,Bernard:2007eh}.

The problem arises because the staggered formulation for lattice
quarks starts with an inherent factor of four in the number of species
\cite{Kogut:1974ag,Susskind:1976jm,Sharatchandra:1981si,Karsten:1980wd}.
These species are sometimes called ``tastes.''  Associated with this
degeneracy is one exact chiral symmetry, which corresponds to a
flavored chiral symmetry amongst the tastes.

As one approaches the continuum limit, the 't Hooft vertex continues
to couple to all tastes.  All of these states will be involved as
intermediate states in the interactions between topological objects.
They give a contribution which is constant as the mass goes to zero,
with a factor of $m^4$ from the determinant being canceled by a factor
of $m^{-4}$ from the sources.

The rooting ``trick'' is an attempt to reduce the theory from one with
four tastes per flavor to only one.  This is done by replacing the
fermion determinant with its fourth root.  However, this process
preserves any symmetries of the determinant, including the one exact
chiral symmetry of the staggered formulation.  This is a foreboding of
inherent problems since the one-flavor theory is not allowed to have
any chiral symmetry.  This symmetry forbids the appearance of the mass
shift that is associated with the 't Hooft vertex of the one flavor
theory.

The main issue with rooting is that, even after the process, four
potential tastes remain in the sources.  The effective vertex will
couple to all of them and be a multi-linear operator of the same order
as it was in the unrooted theory.  Furthermore, it will have a severe
singularity in the massless limit since the rooting reduces the $m^4$
factor from the determinant to simply $m$, while the $m^{-4}$ from the
sources remains.  For the one-flavor theory the issue is particularly
extreme.  In this case the bilinear 't Hooft vertex should be a mass
shift.  However such an effect is forbidden by the exact chiral
symmetry of the rooted formulation.

Another way to see the issue with staggered fermions is via the chiral
rotation in Eq.~(\ref{m5}).  As discussed there, it is essential that
two types of inequivalent mass terms be present.  For staggered
fermions the role of $\gamma_5$ is played by the parity of the site,
i.e. $\pm1$ depending on whether one is on an even or odd lattice
site.  Unfortunately, the exact chiral symmetry of the staggered
formulation gives physics which is completely independent of the angle
$\theta$.  For the unrooted theory this is acceptable since it is
actually a flavored chiral rotation amongst the tastes, with two
rotating one way and two with the opposite effective sign for
$\theta$.  But on rooting this symmetry is preserved, and thus the
regulator cannot be complete.

\section{Summary}
\label{summary}

I have discussed a variety of consequences of the 't Hooft operator.
This is a rather old topic, but many of these consequences remain
poorly understood.  The approach remains the primary route towards
understanding the quantum mechanical loss of the classical axial
$U(1)$ symmetry and the connection of this with the $\eta^\prime$
mass.

The vertex is a direct consequence of the robust nature of exact zero
modes of the Dirac operator.  These modes are tied to the topology of
the gauge fields through the index theorem.  Their stability under
small perturbations of the gauge fields follows from their chiral
nature. 

The form of the vertex exposes interesting discrete symmetries in some
potential models for unification.  Understanding these properties may
be helpful towards finding a non-perturbative regulator for gauge
theories involving chiral couplings to fermions.

This effective interaction ties together and gives a qualitative
understanding of several controversial ideas.  In particular, the
flaws in the rooting trick used in lattice gauge theory become clear
in this context, although they are only beginning to be appreciated.
Also, various attempts to formulate theories beyond the standard model
continue to speculate on a vanishing up quark mass, despite this being
an ill-posed concept.

\section*{Acknowledgments}
This manuscript has been authored under contract number
DE-AC02-98CH10886 with the U.S.~Department of Energy.  Accordingly,
the U.S. Government retains a non-exclusive, royalty-free license to
publish or reproduce the published form of this contribution, or allow
others to do so, for U.S.~Government purposes.



\end{document}